\documentclass[a4paper,fleqn,usenatbib]{mnras}   
\usepackage{xcolor}
\usepackage[T1]{fontenc}
\usepackage{ae,aecompl}
\usepackage{graphicx}	
\usepackage{amsmath}	
\usepackage{amssymb}	
\usepackage{soul}

\def\lesssim{\mathrel{\hbox{\rlap{\hbox{\lower4pt\hbox{$\sim$}}}\hbox{$<$}}}}
\def\gtrsim{\mathrel{\hbox{\rlap{\hbox{\lower4pt\hbox{$\sim$}}}\hbox{$>$}}}}
\newcommand{\ltsima}{$\; \buildrel < \over \sim \;$}
\newcommand{\simlt}{\lower.5ex\hbox{\ltsima}}
\newcommand{\gtsima}{$\; \buildrel > \over \sim \;$}
\newcommand{\simgt}{\lower.5ex\hbox{\gtsima}}

\title[ALMA observations of a proto-GC at z$>6$]{Constraints on the [CII] luminosity of a proto-globular cluster at $z \sim 6$ obtained
with ALMA}
\author[Calura et al.]
{Francesco Calura$^{1}$,
Eros Vanzella$^{1}$,
Stefano Carniani$^{2}$,
Roberto Gilli$^{1}$,
Piero Rosati$^{3}$,
\newauthor 
Massimo Meneghetti$^{1}$,
Rosita Paladino$^{4}$, 
Roberto Decarli$^{1}$,
Marcella Brusa$^{5}$,
\newauthor 
Alessandro Lupi$^{2}$,
Quirino D'Amato$^{4,5}$,
Pietro Bergamini$^{1}$,
Gabriel B. Caminha$^{6}$ 
\\ ~ \\
$^{1}$INAF - Osservatorio di Astrofisica e Scienza dello Spazio, via Gobetti 93/3, 40129 Bologna, Italy\\
$^{2}$Scuola Normale Superiore, Piazza dei Cavalieri 7, I-56126 Pisa, Italy\\
$^{3}$Dipartimento di Fisica e Scienze della Terra, Universit\`a degli Studi di Ferrara, via Saragat 1, I-44122 Ferrara, Italy  \\
$^{4}$INAF - Istituto di Radioastronomia, via Piero Gobetti 101, I-40129 Bologna, Italy\\
$^{5}$Dipartimento di Fisica e Astronomia dell’Universit\`a degli Studi di Bologna, via P. Gobetti 93/2, 40129 Bologna, Italy\\
$^{6}$Kapteyn Astronomical Institute, University of Groningen, Postbus 800, 9700 AV Groningen, The Netherlands
}

\begin{document}

\label{firstpage}

\pagerange{\pageref{firstpage}--\pageref{lastpage}} \pubyear{2019}

\maketitle

\begin{abstract}
We report on ALMA observations of D1, a system at $z\sim 6.15$ 
with stellar mass $M_{*} \sim 10^7 M_{\odot}$ containing globular
cluster (GC) precursors, strongly magnified by the galaxy cluster MACS J0416.1-2403. 
Since the discovery of GC progenitors at high redshift,
ours is the first attempt to probe directly the physical properties of their neutral gas through infrared observations.
A careful analysis of our dataset, performed with a suitable procedure designed to
identify faint narrow lines and which can 
test various possible values for the unknown linewidth value,
allowed us to identify a $4\sigma$ tentative detection of [CII] emission with intrinsic luminosity $L_{\rm [CII]}=(2.9 \pm 1.4)~10^6 L_{\rm \odot}$,
one of the lowest values ever detected at high redshift. 
This study offers a first insight on previously uncharted regions of the 
$L_{\rm [CII]}-SFR$ relation.
Despite large uncertainties affecting our measure of the star formation rate, 
if taken at face value our estimate  
lies more than $\sim 1$ dex below the values observed in local and high redshift systems. 
Our weak detection indicates a deficiency of [CII] emission, possibly ascribed to various explanations, 
such as a low-density gas and/or a strong radiation field caused by intense stellar feedback, and a low metal content. 
From the non-detection in the continuum we derive constraints on the dust mass, with $3-\sigma$ upper limit values as low as $\sim$ a few $10^4~M_{\odot}$,
consistent with the values measured in local metal-poor galaxies. 
\end{abstract}


\section{Introduction}
Observational studies of galaxies at very high redshift 
are crucial to understand how these systems originated and what were the first, fundamental steps of their evolution. 
In the last decade, the Atacama 
Large Millimitre/submillimetre Array (ALMA) enabled a series of 
ground-breaking studies which revolutionized the field of galaxy formation, mostly through 
the detection of strong far-infrared (FIR) lines 
in star-forming galaxies in the early Universe. In particular, the investigation
of the strongest FIR atomic fine structure lines,  such as the
[CII] line at 158$\mu m$ and the [OIII] line at 88 $\mu m$, generally allows one to probe 
the properties of the 
interstellar medium (ISM) in galaxies (e.g. Maiolino et al. 2015; Inoue et al. 2016; 
Laporte et al. 2017; 
Carniani et al. 2017; Matthee et al. 2017; Hashimoto et al. 2018; Carniani et al. 2018a; Bethermin et al. 2020;
Schaerer et al. 2020; Harikane et al. 2020). 
The [CII] line, predominantly originating in photo-dominated regions of actively star-forming galaxies,  
in principle may be useful to trace star forming gas, although with a dependence on metallicity and porosity of the
ISM (De Looze et al. 2014). This line is generally used to probe the physical conditions of the  
ISM in galaxies at $z>4.5$, i. e. in an epoch which marks the transition from 
cosmic reionization, which presumably was completed at $z \sim6$, to the appearence of dust- and metal-rich galaxies,
already observed at comparable redshifts (e, g., Nagao et al. 2012;  Marrone et al. 2018). 

Most infrared/sumbmillimeter observations carried on so far in high redshift objects were performed by pointing instruments on bright sources, i. e. in 
actively star forming galaxies such as Lyman-$\alpha$ emitters, in  quasar/AGN host galaxies and submillimeter-detected starbursts 
(Venemans et al. 2012; Wang et al. 2013; Willott et al. 2013; Riechers et al. 2013, Decarli et al. 2018;
for a review, see Hodge \& da Cunha 2020) and fewer detections in normally star-forming galaxies,
in some cases selected with the Lyman Break technique
(Willott et al. 2015; Bakx et al. 2020). 
So far, very little attention has been devoted 
to the search for [CII] emission in particularly faint systems, for obvious technical reasons.
However, the study of [CII] emission in faint objects, which presumably are associated to low-surface brightness objects such
as extended star-forming complexes of dwarf galaxies is particularly interesting,
as in the domain of these systems the [CII]-star formation rate (SFR) relation found in local galaxies seems to break (e. g., De Looze et al. 2014).
Moreover, the study of very faint systems towards the reionization epoch is of particular interest 
also because some of these objects have been proposed as important contributors to reionization itself, perhaps in the form
of dwarf galaxies (e. g., Bouwens et al. 2012)  
or more isolated compact, massive star clusters (Ricotti 2002). In fact, in these systems the leak of ionizing photons is
thought to be particularly favoured, owing to low HI column densities, or to extended, ionized cavities created by concentrations 
of massive stars (Vanzella et al. 2020a).\\
A natural help in observational studies of particularly faint objects is provided by strong gravitational lensing.
In the last few years, much progress in this field has been driven 
by deep observations of massive galaxy clusters, carried out in the context of large Hubble Space Telescope programmes, in particular the Hubble
Frontier Fields (HFF) survey (Lotz et al. 2017). 
Robust lens models of galaxy clusters are built thanks to the identification of large 
numbers of multiply lensed sources, which can span a large redshift range (Meneghetti et al., 2017; Bergamini et al. 2019).
Thanks to strong lensing effects produced by massive clusters of galaxies located along the line of sight, sources can be magnified by large
factors (ranging from $\mu$ of the order of a few up to $\sim~100$), allowing faint, compact objects
to be studied with very high spatial resolution and signal-to-noise ratios (SNR) and in some cases,
to probe their structural parameters down to scales of a few $\sim 10$ pc.
In this context, the determination of the redshift of the images is a key problem. In many cases, photometric redshift estimates are accessible,
but a safe determination of the redshift generally has to rely upon a spectroscopic detection, achievable only with deep observations.
In this regard, in recent times considerable, further progress was possible thanks to the significant use 
of powerful instruments such as the Multi Unit Spectroscopic Explorer (MUSE, Bacon et al 2010) mounted on the VLT, which 
has enabled the spectroscopic confirmation of hundreds of multiple images at high redshift ($z>3$, e. g., Caminha et al. 2017; Lagattuta et al. 2019). 
This has enhanced the production of highly accurate lens models, significantly mitigating systematic uncertainties
in the computation of magnification maps in lensed fields. This recent progress has allowed one to determine absolute physical
quantities such as luminosities, sizes, stellar mass values and star formation rates of new objects,
which before were impossible to study in non-lensed fields. \\
One of the most important results achieved with Deep MUSE pointings in strongly lensed fields was the first discovery of the precursors of
globular clusters (GC). Vanzella et al. (2017a) presented a detailed study of a few very faint
systems detected behind two HFF clusters between  $z\sim~3$ and $z=6.1$, for which accurate measures of
the radial surface brightness profiles, sizes, spectral energy distributions, stellar masses and star formation rate (SFR) values were obtained.
The conclusion that some of these systems might qualify as GC precursors stems from the 
study of their stellar mass values (between $10^6$ and $10^7 M_{\odot}$) and sizes
(of the order of a few 10 pc, Vanzella et al. 2017a; Vanzella et al. 2017b, Bouwens et al. 2017; Kikuchihara et al. 2020)
with stellar surface density values compatible with the expectations of 
some GC formation scenarios (e. g., D'Ercole et al. 2008, Calura et al. 2019). 
The most striking example of a GC precursor was found as a component of an extended star-forming 
complex at $z=6.1$, with a total stellar mass of $\sim~2 \times 10^7~M_{\odot}$ and which 
contains a compact, nucleated, very dense region at its centre (Vanzella et al. 2019). 
The nuclear component has a stellar mass of $\sim10^6~M_{\odot}$, young stellar populations ($<10$ Myr) and for its 
effective radius an upper limit of 13 pc was derived. Other individual star-forming knots are present in a widely diffused star
forming complex at $z=6.1$, with intrinsic UV magnitudes between 31 and 32 and sizes between 10 and 50 pc (Vanzella et al. 2019). \\
In this paper we report on ALMA band 6 observations of the strongly magnified, central region 
of this star forming complex, aimed at the detection of [CII] emission.  
To our knowledge, this study is one of the first in which such a faint object (with intrinsic UV magnitude $m=29.6$, Vanzella et al. 2019)
and at such a high redshift has ever been targeted with ALMA (see also Knudsen et al. 2016).\\
Our paper is organised as follows. In Section 2 we describe our target, our observations and the methods used for data reduction.
In Section 3 we discuss our results and finally we present our conclusions in Section 4. \\
Throughout this paper we adopt a Salpeter (1955) stellar initial mass function (IMF) and concordance cosmology with $\Omega_m=0.3$, $\Omega_\Lambda=0.7$, and $h=0.7$.
At the redshift of our source, this implies a scale of 5.63 kpc/\arcsec  and a luminosity distance $D_{\rm L}=5.94\times10^4$ Mpc (Wright et al. 2006).

\section{Target description and observations} \label{sec_obs}
The target, shown in Fig. 1, was initially identified by Caminha et al. 2017
and subsequently analysed by Vanzella et la. (2017a, hereafter V17),
where several compact, strongly lensed objects at $z>3$ were analysed.  
The target was first detected by means of a two hours MUSE integration,
which revealed three multiple images of the same Lyman-$\alpha$ (Ly$\alpha$) nebula.
We centered our ALMA pointing on this system (Fig. 1, left panel),  
a superdense star-forming region at $z=6.1$ composed of two compact sub-systems, D1 and T1 (shown in the top-right inset of Fig. 1), detected by the Hubble Space Telescope (HST) in the WFC3/F105W and ACS/F814W bands. 
The system including D1 and T1 is part of an extended ($\sim$ 50\arcsec across) Ly$\alpha$ arc detected in the Hubble Frontier Field galaxy cluster MACS J0416. 
As discussed in V17, in several cases the gravitational lens (in general a galaxy cluster) causes a significant stretching of the source,  
allowing one to probe very small sizes, sometimes of the order of a few 10 pc or less, and to spatially resolve substructures, 
such as multiple clusters  
in extended star forming complexes. The subject of this work is one of the best-studied cases, in which separate stellar components 
are revealed from UV oservations, and stellar masses, star formation rates as well as half-light radii can be derived
for several single sub-components. 

The redshift computed from the Ly$\alpha$ line in deep MUSE observations is $z=6.149 \pm 0.003$ (Vanzella et al. 2020b).

\subsection{D1 and T1 as infant globular clusters} 
The D1-T1 system was analysed in detail in Vanzella et al. (2019, V19 hereinafter),
where an accurate model of the gravitational lens was presented, 
able to reproduce with great precision 
the positions of several spectroscopically confirmed multiple images at $3\le z\le 6.5$ (Caminha et al. 2017; Bergamini et al. 2020). 

The UV emission is generated by extremely faint systems (with intrinsic magnitudes $m_{\rm 1500}=28-33$), presenting stellar masses 
of a few $10^{6}$~M$_{\rm \odot}$ and consistent with ages of 1-10 Myr (V17) obtained from the analysis of
their spectral energy distribution (SED, see Sect.~\ref{sec_sfr}) . \\
The sources are located in a region of high magnification ($\mu>15$, see Fig. 1).  
The systems D1 and T1, separated by 1.7\arcsec, could be spatially resolved along the tangential direction
(further details can be found in V17, V19).    
From the light profiles observed in the $F105W$ HST band 
(which probes the rest-frame 1500 \AA) and by means of GALFIT modelling,
it was possible to obtain an estimate of the size of T1, D1 and other star-forming knots in their proximity,
with values spanning  the  interval  13 pc$\le R_e <$50 pc (V19). 
In V19 the main features of the target of our ALMA observations were presented, summarised in Table~\ref{tab1}.

\begin{table*}

\small
  \caption{Summary of the main physical properties of D1 as presented in V19.
  First column: redshift. Second column: total magnification. Third column: stellar mass. Fourth column: age of the system.
  Fifth column: SFR. Sixth column: apparent magnitude. Seventh column: absolute UV magnitude.
  Eighth column: effective radius. }
\begin{tabular}{cccccccc}

\hline
$z^a$ & $\mu$ & $M_*^b$ & Age$^c$     & SFR$^{b,c,d}$      &  $m(1500\AA)$$^b$ & $M_{\rm UV}^b$ & $R_{\rm e}^b$\\
     &       &  $(M_{\rm \odot})$ & (Myr)& $(M_{\rm \odot}/yr)$ &          &         &(pc) \\ 
\hline
z=6.149{\tiny$\pm0.003$} & 17.4 $\pm~6$ & $2.2_{-0.06}^{+1.1}\times~10^7$ & 1.4$_{-0.4~(-0.4)}^{+1.6~(+98.6)}$  & 15.8$_{-8.0~(-15.4)}^{ +17.8~(+61.8)}$ \small & 29.6$\pm0.2$ & -17.1$\pm0.2$  & 44 $\pm19$ \\
\hline\\
\end{tabular}
\begin{tabular}{l}
$^a$ Computed from the  MUSE Ly$\alpha$ detection, using the air-based Ly$\alpha$ wavelength value of $1215.136 \AA$. \\
$^b$ All quantities are intrinsic, i. e. corrected for magnification due to gravitational lensing.\\
$^c$ For each measurement we report the $1 \sigma$ ($3 \sigma$) uncertainties.\\
$^d$ The lower $3 \sigma$ errorbar of the SFR is computed considering the available constraints from the Ly$\alpha$.\\
detection, whereas the upper errorbar is computed at 3$\sigma$.\\
\end{tabular}
\label{tab1}
\end{table*}

With an intrinsic UV magnitude of 31.3, T1 is one of the faintest spectroscopically confirmed star-forming
objects ever identified at such a high redshift. 
Overall, in available HST images D1 appears considerably more extended than T1 and
it is characterised by a nucleated star-forming region surrounded by a diffuse component.  
Despite the large magnification, the spatial distribution of the central emission appears circular and symmetric,
which implies a small size of the innermost component. 
By means of SED fitting techniques and considering a plausible value for the magnification factor,
V19 estimated for D1 a total stellar mass of $\sim2 \times 10^7~M_{\odot}$, a star formation rate (SFR) of  $16~M_{\odot}/yr$ 
and an effective radius of 44 pc, whereas for its nucleated central component a stellar mass of $\sim10^6~M_{\odot}$ and effective radius
$<13$ pc. 
Due to the peculiar features of these two objects, i.e. their  
extremely small physical sizes, absolute magnitudes $M_{\rm UV}\ge-16$ and their stellar mass values, D1 and T1 
can be regarded as two of the best cases for globular cluster  
precursors ever discovered so far at high redshift, and among the densest star-forming objects ever found.\\
Overall, the target of our observation appears as a richly structured star-forming complex which includes
a variety of compact sources, and for such features is a unique laboratory to study the physical properties of
very dense star-forming objects. 
With the present observations we targeted the central, most magnified image of the Ly$\alpha$ emitting region.
In the remainder of the text, for brevity we will refer to the D1-T1 system with 'D1', i .e. with the name of the system
which dominates the star-forming complex in terms of UV luminosity.\\
The primary aim of our ALMA program was to detect via the [CII] emission line at 158 $\mu m$ (rest-frame) the possible presence of cold gas in the systems. 
The one presented in this paper is the first observational study performed at submm wavelengths of such a 
rich and unique environment, representative of the one in which the oldest and most metal poor globular clusters might have originated.

\begin{table*}
\small
  \caption{Summary of the physical properties of D1 derived in the present study from our ALMA observations. 
  (1): observed frequency of the line. (2): redshift calculated from the [CII] detection. 
(3): line sensitivity. 
(4): continuum sensitivity.
(5): linewidth (FWHM). 
  (6): continuum flux limit at $3~\sigma$.
   (7): [CII] peak flux density with error. (8): [CII] line-integrated density flux with error .  
  (9): [CII] luminosity of D1 with error. (10): upper limit on the dust mass of D1 assuming dust properties as in Ouchi et al. (2013). 
  (11): upper limit on the dust mass of D1 in the case of I Zw18-like dust.}
\begin{tabular}{ccccccccccc}
\hline
 (1) & (2) & (3) & (4) & (5) & (6) & (7) & (8) & (9) & (10) & (11)\\ 
\hline
$\nu_{\rm obs}$ & $z_{\rm [CII]}$    & $\sigma_{\rm line}$$^a$  & $\sigma_{\rm cont}$ & FWHM   & $S_{\rm cont}$  & $S^{peak}_{\rm \rm [CII]}$   & $S_{\rm [CII]}~\Delta v^b$  & $L_{\rm [CII]}$$^{b,c}$     & $M_{\rm  dust}$$^{~c,d}$  &  $M^{IZ18}_{\rm  dust}$$^{~c,e}$  \\

(GHz)  &    & {\tiny$(\mu~Jy~beam^{-1})$} &  {\tiny$(\mu~Jy~beam^{-1})$}  & {\tiny($km~s^{-1}$)} &  {\tiny($\mu~Jy$)} & {\tiny($\mu~Jy$)} &   {\tiny$(Jy~km~s^{-1})$}  & $(10^6 L_{\odot})$ &  $(10^5 M_{\odot})$ & $(10^5 M_{\odot})$ \\
\hline
265.83{\tiny$\pm$0.02} & 6.1495{\tiny$\pm$0.0006}    &106       &     9.6      &   110$\pm$50        & $<$28  &  446$\pm$106  & 0.05$\pm$0.02  &  $2.9 \pm 1.4$ & $<3.2$ & $<0.43$ \\
\hline
\end{tabular}
\begin{tabular}{l}
Notes:\\
$^a$Estimated in a spectral channel of 110 km/s. \\
$^b$The integrated flux density and luminosity are estimated from a circular aperture of radius 1\arcsec. We compute the flux uncertainties 
rescaling the noise \\by $\sim\sqrt{N}=2$, namely the square root of the number of independent beams in the circular region.\\
$^c$All quantities are intrinsic, i. e. corrected for magnification due to gravitational lensing, and assuming $\mu=17.4$. \\
$^d$Computed assuming a spectral index value  $\beta=1.5$ and dust temperature $T=40$ K (Ouchi et al. 2013).\\
$^e$Computed assuming a spectral index value $\beta=2.0$ and dust temperature $T=70$ K (Hunt et al. 2014).\\
\end{tabular} 
\label{tab2}
\end{table*}

\subsection{Derivation of the Star Formation Rate and other physical properties of D1}
\label{sec_sfr}
Physical properties of D1 such as SFR, stellar mass and age 
were derived by means of SED-fitting techniques and based on the Astrodeep photometry (Merlin et al. 2016).
The age and SFR values estimated for D1 are reported in Tab.~\ref{tab1}, along with their associated $1\sigma$ and $3\sigma$ uncertainties. 
More details on their derivation can be found in V17 and V19.

The fits were performed with a set of standard templates (Bruzual \& Charlot 2003) which include also 
nebular continuum and emission lines (Castellano et al. 2016).
Despite the extremely faint intrinsic magnitude of D1, 
the multiband photometry is robust thanks to long HST exposure times matched to strong lensing magnification and 
a SNR $>$ 20 was achieved in all the HST/WFC3 bands.
In general, due to the lack of constraints at optical rest-fame wavelengths, 
the SED-fitting analysis shows degenerate solutions for 
stellar mass, age, and star formation rate values (V17). 
However, the Ly$\alpha$ flux measured from MUSE spectroscopy allows us to strengthen the constraints
on both SFR and age.
The rest-frame equivalent width (EW) of the Ly$\alpha$ line
was computed by integrating the Ly$\alpha$ flux over different apertures that
include D1 and a more extended region which embraces a large portion of the Ly$\alpha$ arc.
In the most conservative case, assuming negligible dust contribution
the resulting rest-frame EW is $60 \pm 8$ \AA~which, if compared  with the evolution of
the same quantity extracted from stellar population synthesis (Schaerer 2002)  
suggests an age of D1 younger than 100 Myr assuming constant SFR (and of $\sim 5$ Myr assuming instantaneous star formation).

As discussed in V19, the intrinsic SFR derived from the SED-fitting at 3$\sigma$,
including the constraints from the Ly$\alpha$ emission, ranges between 0.4 M$_{\odot}~yr^{-1}$ and $\sim 78$ M$_{\odot}~yr^{-1}$, 
with 15.8  M$_{\odot}~yr^{-1}$ as best value. 
The very blue ultraviolet spectral slope ($\beta_{\rm  UV}=-2.5~\pm~0.10$, where the UV flux is $F_\lambda \sim \lambda^\beta_{\rm  UV}$, V19) 
of D1 confirms its young age and suggests very little dust attenuation.

\subsection{Observations and data reduction}
\label{sec_obs}
In our program, ALMA observations of the central,
strongly magnified central image of a giant Ly$\alpha$ nebula containing 
D1 were performed to detect the [CII] emission line and continuum emission.
Our aim was to detect the $^2P_{3/2}-^2P_{1/2}$ [CII] transition 
at 157.74 $\mu m$ (1900 GHz) rest-frame at a fequency of 266.05 GHz in the observer frame. 
The system was observed with ALMA in band 6 during Cycle 6 in the program 2018.1.00781.S. 
The observations consisted of 3 hours of on-source integration, 
carried on with the C43-3 nominal configuration in December 2018 and April 2019 in three
separate execution blocks.
In the three blocks 
the shortest baseline was of 15 m and the longest was between 483 m and 649 m 
(with nominal maximum angular resolution between 0.5\arcsec and 0.6\arcsec),  
with maximum recoverable scale between 6\arcsec and 7\arcsec.
Considering the magnification values obtained from the lens model for MACS J0416 at the coordinates of D1, 
a spatial full-width half-maximum resolution of 0.5\arcsec is suitable for probing a delensed spatial extent of $\sim$200 pc.\\
Four different spectral windows (SPWs) were chosen, one at central frequency 266 GHz and
with a spectral resolution of 1129 kHz (corrisponding to a $\sim 1$ km/s velocity resolution), 
and three at the lower-frequency side of the [CII] line, at 
central frequency values 264 GHz, 251 GHz and 249 GHz with 7.8 MHz resolution ($\sim~9$ km/s),
all used for continuum measures and calibration purposes.
Given the peculiar nature of the source, the velocity width of the line was unknown a priori.
In principle, a maximum  spectral window resolution of ~1 km/s, of the order of the velocity dispersion of star
clusters, was chosen in order to identify the line also in case it was particularly narrow. \\
The dataset has been calibrated by using standard  scripts for
ALMA data reduction and in particular with the CASA software (version: 5.4.0-70; see McMullin et al. 2007).\\
No continuum subtraction was detected in the field of view,
therefore no continuum subtraction has been performed on the dataset. \\
Line emission and continuum maps from the four SPWs were obtained by means of the CASA {\it clean} task 
and adopting a natural weighting.

The continuum image and line datacube have an angular resolution of $1.03\arcsec\times0.63\arcsec$~(PA\footnote{Position angle (North through East).}~$=-86^\circ$)
and $0.99\arcsec\times0.60\arcsec$~(PA~$=-85^\circ$), respectively. 

A summary of the main results of the observational program obtained after the reduction procedure are reported in Table~\ref{tab2}.  

\begin{figure*}
  \includegraphics[height=10cm, width=13cm]{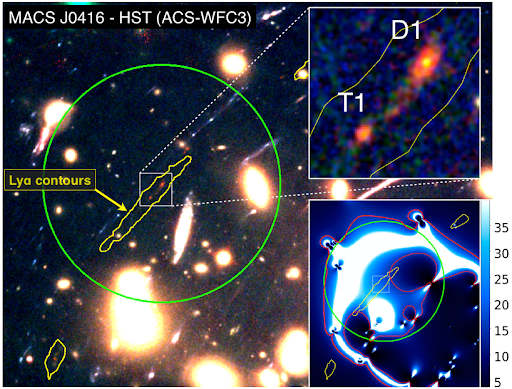}
  \caption{HST composite colour (red=F105W, green=F814W, blue=F606W) 35\arcsec$\times$35\arcsec 
  image of the Lyman-$\alpha$ arc in the galaxy cluster MACS J0416 
(traced by the yellow contours, plotted at 1-$\sigma$) at $z=6.145$ observed in our ALMA program. 
    Our target contains two very compact star-forming systems, D1 and T1, plotted here with the  22\arcsec ALMA field of view (green circle).
    The 3\arcsec$\times$3\arcsec inset in the top-right shows  
    a zoom-in of the region containing D1 and T1. Bottom inset:  magnification map of the entire area. 
    The red countour encloses a region with magnification $\mu=15$.}
    \label{fig_obs}
\end{figure*}

\section{Results}
Although this is not the first attempt to
study [CII] emission in a lensed field (see, e.g. Schaerer et al. 2015; Bradac et al. 2017
Dessauges-Zavadsky et al. 2017, Gonzalez-Lopez et al. 2017), 
it is one of the very first to pursue such a task in a system as faint as D1. 
A thorough investigation of the ALMA data cube performed by means of an automatic procedure
led us to identify a tentative $\sim$4$\sigma$ detection 
in the [CII] maps, which will be detailed below. This led us to obtain an estimate of the [CII] luminosity
of D1. We will use the SFR derived in V19 and we will study D1 in the $L_{\rm [CII]}$ -SFR relation, a useful probe of 
the properties of the ISM in star-forming galaxies.

\begin{figure*}
\includegraphics[height=6cm, width=15cm]{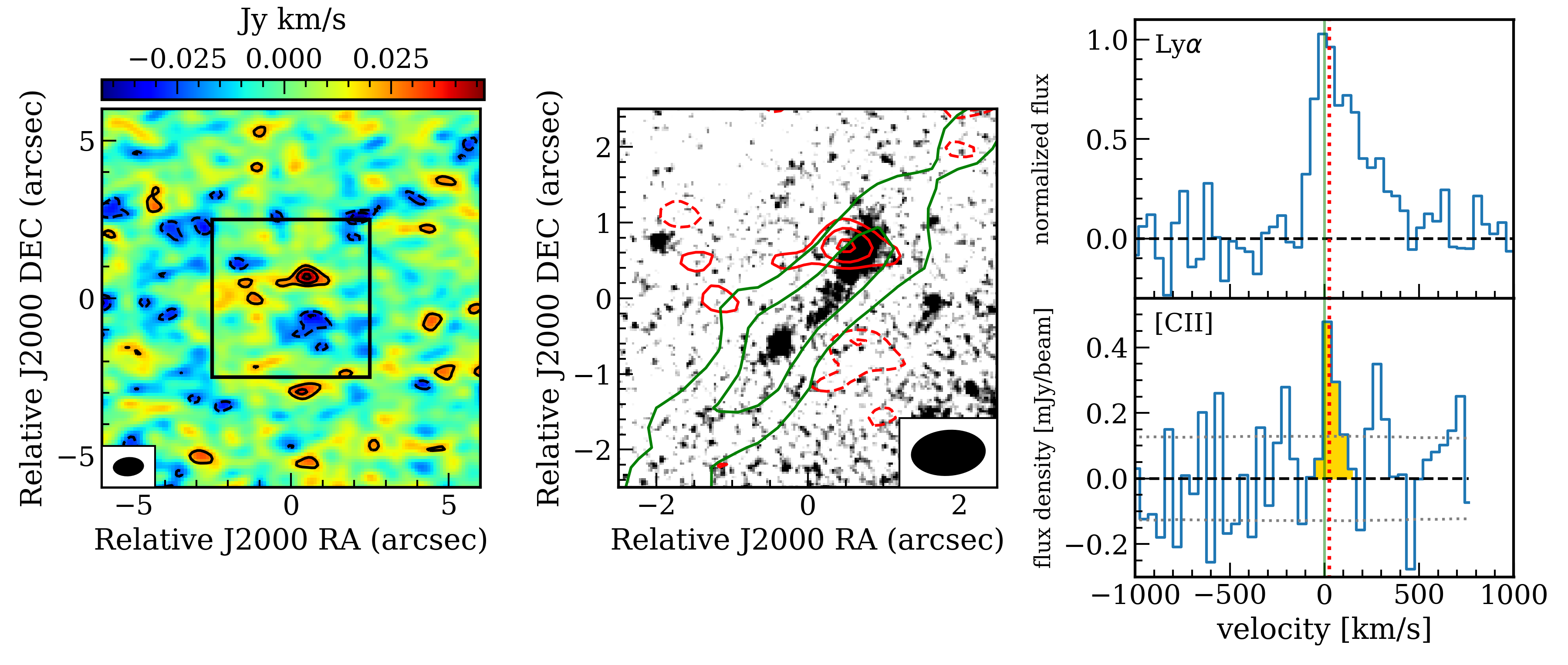}
\caption{This figure illustrates the main features of the tentative [CII] detection discussed in Sect.~\ref{sec_ten}.
The left panel shows a 12”$\times$12” map of the [CII] emission,
enclosed in the black square.
The map was obtained by collapsing the datacube in frequency around the candidate line.  
The black solid contours show the 2$\sigma$, 3$\sigma$, and 4$\sigma$ levels, where $\sigma = 0.011$~Jy km/s. 
The ALMA beam is indicated in the in bottom-left inset of this panel
In the middle panel, the background black-and-white image shows a HST map of a 5”$\times$5\arcsec region containing D1 and T1, whereas
the red and green solid contours show the [CII]  and the Ly$\alpha$ emission, respectively. 
In the right panel, the top figure shows the Ly$\alpha$ profile (V19), whereas 
the right figure  shows the [CII] spectrum, as extracted from the brightest pixel of the [CII] map and computed with a spectral rebinning of 44 km/s.
The thick, black dashed horizontal line indicates the zero-flux level, whereas the gray dotted lines indicate the $1\sigma$ error.
The red dashed lines mark the velocity offset of the [CII] with respect to the Ly$\alpha$ detection,
whose position is indicated by the green solid line. }
\label{fig_tent}
\end{figure*}

\subsection{A tentative [CII] detection}
\label{sec_ten}
The raw data collected in the SPW at central frequency 266.054 GHz were carefully analysed with a method 
targeted to identify weak, narrow emission lines in ALMA datacubes.\\ 
An automatic method was used to perform a blind search with a 5”$\times$5” aperture,
which was placed at the centre of the field of view and through the entire SPW.
An automatic search throughout the entire ALMA cube was performed,
aiming for a feature at various spatial positions and spanning 
several possible values for the linewidth $\Delta v$, which is unkown for an object like the one discussed in this paper.\\

Detailed predictions regarding the features of atomic lines such as [CII] do exist as
derived from hydrodynamical, zoom-in simulations in a cosmological context of large galaxies (Pallottini et al. 2017) or dwarfs 
(Lupi \& Bovino 2020), but no estimate is available at the scale of a system with stellar mass of the order of $10^7~M_{\odot}$. 
As an initial guess, it is perhaps reasonable to assume a linewidth of the order of the velocity dispersion $\sigma_v$ of 
local stellar systems of comparable stellar mass,  i. e. of $M_{\rm  *}\sim10^7~M_{\odot}$ dwarf galaxies, 
which typically show values for $\sigma_v$ of the order of a few ten km/s (e. g., Walker et al. 2007; Cody et al. 2009).
Based on these arguments, for the linewidth value the range between  $\Delta v \sim 10~km/s$ and  $\Delta v \sim 200~km/s$ was explored. 

After a careful inspection of the identifications, a tentative [CII] detection was found, illustrated by Fig~\ref{fig_tent}. 
The left panel of Fig.~\ref{fig_tent} shows
a coloured  
12\arcsec$\times$12\arcsec~[CII] map, where the black contours highlight the detection at $2-4 \sigma$. 
In the middle panel we show a black-and-white HST map where D1 is visible, reported 
with the tentative ALMA detection, represented by the solid red countours. 

The emission peak has SNR$=4.2$, and the spatial offset between the centre of the 
UV detection of D1 and [CII] is $\sim 0.15$\arcsec-0.2\arcsec.
Given the major-axis of the ALMA beam, the spatial offset lies within the astrometric
uncertainty of $\sim0.2\arcsec$.  

By performing a Gaussian fitting on the [CII] spectra, we estimated the main properties of the emission line. 
The linewidth is 110 $\pm$ 50 km/s and the velocity offset  with respect
to the Ly$\alpha$ is +21 $\pm$ 128 km/s.

The integrated flux density of 0.05$\pm$0.02 Jy km/s  has been estimated
from a circular aperture of radius 1\arcsec and centered at the location of the emission peak of
the [CII] map (see left panel of Fig.~\ref{fig_tent}). 
The error on the integrated  flux density has
been rescaled by the square root of the number of beams contained within the aperture (N=4). 

A note is in order concerning the ALMA beam (reported in the bottom-left insert of the first panel), which 
is strongly tilted with respect to the detected UV emission and it has an
orientation very similar to the tentative ALMA detection. 
The shape and orientation of the beam could severely limit 
the possibility of having a clearer detection if an extended emission was present and inclined  with respect to
the major axis of the ALMA beam, e. g. parallel to the axis of the UV emission of D1.\\
The feature visible in Fig.~\ref{fig_tent} and associated to the tentative detection was the only one found
by means of this method with such a high statistical significance. 
Other few weak hints of a detection are present within 2.5\arcsec~from the centre of D1, 
obtained with different spectral binning, 
which remain marginal as they generally present SNR$ \le 3$.
It is worth stressing that a SNR$=4.2$ does not fully prove this to be 
a real detection and not a strong noise fluctuation.
Two concrete reasons for regarding this feature as a tentative detection
are its position and velocity values, consistent with the estimates 
suggested by other measures at different wavelengths. 
We attempted to taper the image to probe more extended emitting regions.
The tapering is generally recommended for weak, extended sources and sensitive to the geometry of the source
(Carniani et al. 2020), 
therefore we performed a few attempts testing a range of possible
shapes for the emitting region,  
but without obtaining any subtantial SNR improvement. \\
The spectrum extracted by the brightest pixel is shown in the right panel of Fig. ~\ref{fig_tent} (see its caption for further details),
which yields a systemic redshift value $z=6.145$. \\
The robustness of our detection is further investigated in Appendix~\ref{sec_mock}, where the effects of
the geometrical distribution of the emitting region on the estimated SNR are evaluated.

\begin{figure*}
  \includegraphics[height=10cm, width=12cm]{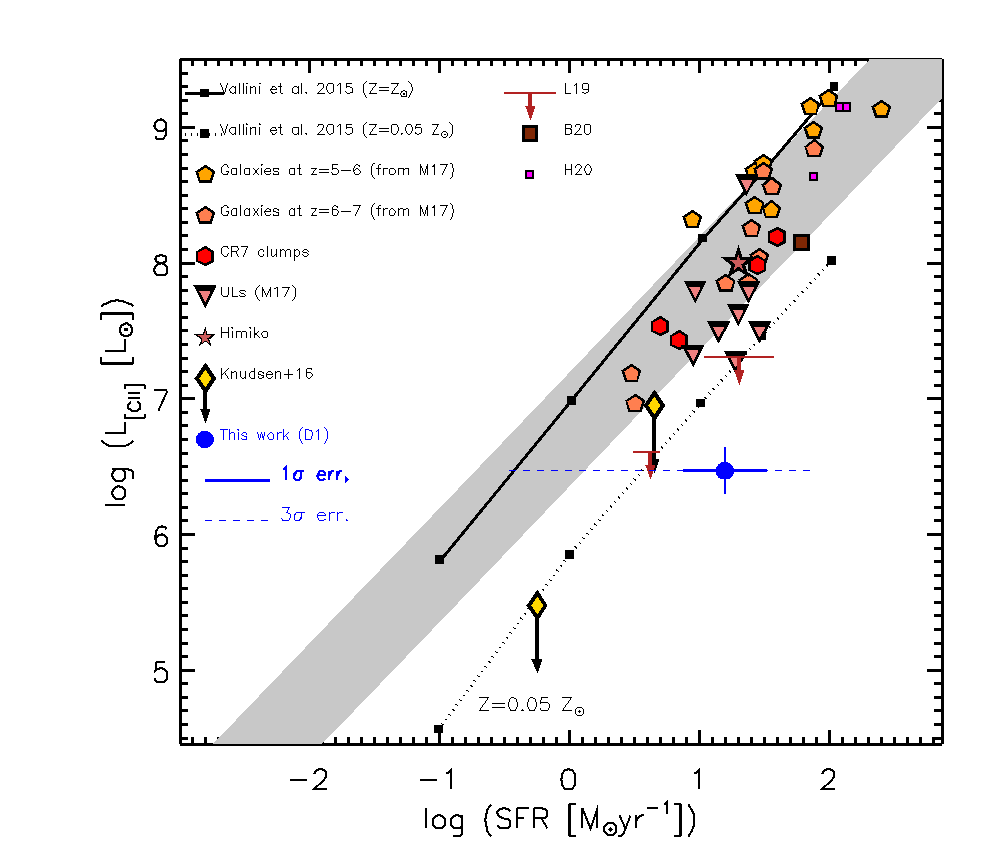}
  \caption{Observed relation between [CII] luminosity and star formation rate (in general determined from UV and FIR observations)
  in local and distant galaxies as obtained 
  in the present study, compared with data from the literature. 
  The tentative detection value presented in Sect.~\ref{sec_ten} is the blue solid circle, reported with the 
  $1\sigma$ (blue thick solid line) and $3\sigma$ (blue thin dashed line) error bars.  
  As for the data from other authors, where required the SFR values have been corrected for the IMF, which in the present work is the one of Salpeter (1955).
  The orange and dark yellow pentagons and the red hexagon are galaxies between $z=6$ and $z=7$, between $z=5$ and $z=6$ and clumps of the CR7 objeect,
  respectively, all 
  from Matthee et al. (2017).
  The light-red inverted triangles are upper limits from Matthee et al. (2017). The yellow diamonds are ALMA observations of
  lensed systems at $z>6$ (Knudsen et al. 2016); in this case, the SFR values are the ones from UV-optical observatons. The pale red star is the estimate 
  for the Himiko galaxy by Carniani et al. (2018b).
   The dark red lines with downpointing arrows are upper limits measured in two systems at $z>8$ by Laporte et al. (2019).
  The large brown square is the recent estimate derived by Bakx et al. (2020) in a LBG at z=8.3. 
The small magenta squares are estimates for galaxies at $z>6$  from Harikane et al. (2020).  
  The grey area represents the local relation estimated by De Looze et al. (2014) computed for their entire
  galaxy sample, reported with its $1-\sigma$ dispersion. The dotted and solid black lines are the model of Vallini et al. (2015) computed for a
  $Z=0.05~Z_{\odot}$ and $Z=Z_{\odot}$ metallicity values, respectively. }
  \label{fig_cii}
\end{figure*}

\subsection{The [CII] luminosity of D1}
\label{sec_lcii}

The [CII]158$\mu m$ fine structure line is  
a very efficient and dominating coolant for neutral gas, originating mostly in photo-dominated regions and 
arising from the transition of singly ionized carbon atoms (C+) from the
$^2P_{3/2}$ to the $^2P_{1/2}$ state, and among the strongest emission line in the far-infrared. 
A significant fraction of gaseous carbon is expected to be in the form of C+ under a wide range of physical conditions, including
ionized, atomic, and molecular phases of the ISM (Goldsmith et al. 2012). 
For a reasonable range of gas temperature values (between $10^3$  and $10^4$ K), this transition occurs in a gas with a critical density  
of the order of $\sim 10^3~cm^{-3}$ (Goldsmith et al. 2012). Hence, the fact that 
[CII] emission is able to trace cold, neutral gas is reasonably well ascertained in the literature.
The main question is whether [CII] emission traces cold star-forming gas or not. To address this fundamental issue, several studies
have been dedicated to the relation between $L_{\rm [CII]}$ and SFR in local and distant galaxies.
In Fig.~\ref{fig_cii} we show the $L_{\rm [CII]}$-SFR as determined in both local and distant galaxies by various authors,
along with the [CII] luminosity derived in the present study with ALMA for D1.\\ 
The [CII] luminosity of D1, expressed in solar luminosities $L_{\odot}$,
has been computed from the flux density  value $S^{peak}_{\rm [CII]}$ reported in Tab.~\ref{tab2} by means 
of the following formula:
\begin{equation}
L_{ \rm [CII],D1}=  1.04 \times 10^{-3} \times (S_{\rm [CII]}~\Delta v)~D^2_L~\nu_{\rm  obs}~L_{\odot},
\label{eq_lcii}
\end{equation}
(Carilli \& Walter 2013).
The quantity $D_L$ is the luminosity distance expressed in Mpc,
whereas $\nu_{\rm  obs}=266$ GHz is the rest frequency of the line. 

The [CII] luminosity is $L_{\rm [CII], D1}=(2.9 \pm 1.4)~10^6 L_{\odot}$, corrected for lensing and
shown by the blue solid circle in figure~\ref{fig_cii}.  
In Fig.~\ref{fig_cii}, our value is compared to  a collection of other estimates from the literature 
as obtained in both local and high-$z$ galaxies. 
Despite the large uncertainty on the SFR value, 
our estimate occupies a region of the plot where extremely few previous measures are present, in particular 
regarding high-redshift systems. 
A discussion on the comparison of our estimate with previous measures
of $L_{\rm [CII]}$ from the literature 
is postponed to Sect.~\ref{sec_disc}.

\subsection{Discussion}
\label{sec_disc}

Observationally, [CII] emission is detected primarily in star forming galaxies, 
however its reliability as a SFR tracer,  as well as the existence of
a unique relation between $SFR$ and $L_{\rm [CII]}$ and its evolution are still largely debated issues.
In the remainder of this Section, we extend the comparison of our results on the [CII] luminosity of D1 
with previous estimates, we discuss the implications of our results in the context of studies of
young star clusters at both low  and high redshift, and finally 
we discuss further constraints on the dust content of D1 from
the non-detection of the continuum.

\subsubsection{On the $L_{\rm [CII]}$-SFR relation at high-z}
In the Herschel Dwarf Galaxy Survey,
De Looze et al. (2014) observed a sample  of spatially  resolved low-metallicity dwarf galaxies
in the local Universe. 
Their study showed a tight (with a $\sim$ 0.3-0.4 dex $1~\sigma$ dispersion)  
relation between $L_{\rm [CII]}$ and SFR and in a variety 
of physical conditions, which extends to a broad sample of galaxies of various types and metallicities from the literature. 
The general result is that the [CII] emission line is to be regarded as a reliable SFR tracer in starbursts or in average star forming galaxies. 
One of the conclusions of their analysis, relevant for the present study, 
was that the [CII] line is less of a reliable SFR tracer in galaxies of low metal abundance.
A few local low-metallicity galaxies showed an offset in the $L_{\rm [CII]}$-SFR relation compared to more metal-rich galaxies.  
This was confirmed by a sensible increase of the scatter in the $L_{\rm [CII]}$-SFR relation towards low metallicities and as a
function of other parameters such as  
warmer dust temperatures and increasing filling factors of diffuse, highly ionized gas.
In these conditions, other cooling lines 
are more dominant with respect to [CII] (such as, e. g., [O I], De Looze et al. 2014). \\
Through the detection of [CII] emission in star-forming galaxies at high redshift, 
it was possible to achieve a deep characterisation 
of the cold ISM of several galaxies at the end of reionization (Hodge \& Da Cunha 2020 and references therein). 
The significant progress in studies of the $L_{\rm [CII]}$-SFR relation at high redshift, in great part driven by ALMA, concerned mostly
bright systems.\\ 
Extended studies of this relation in high redshift galaxies were reported in various works, without reaching a clear consensus
on whether the $L_{\rm [CII]}$ - SFR relation is redshift-dependent or not. 
Some of these results reported on several cases of CII-underluminous galaxies with respect to the local $L_{\rm [CII]}$ - SFR relation, sometimes 
indicating a larger scatter 
(Maiolino et al. 2015; Willott et al. 2015; Pentericci et al. 2016; Matthee et al. 2017, Carniani  et  al.  2018a),
including also lensed systems (Bradac et al. 2017).
Other subsequent studies, sometimes based on the re-analysis of extant datasets, found also in high-redshift systems (i.e. a $z\sim 6-7$)
a $L_{\rm [CII]}$ - SFR consistent with the local one (e. g., Matthee et al. 2019; Carniani et al. 2020), at least for the most luminous galaxies,
whereas in faint systems deviations from the local relation were still found.
Again, Harikane et al. (2020) found that at a given SFR, galaxies at $z>5$ are weaker [CII]-emitters than local ones, 
and that this deficit is dependent on their Ly$-\alpha$ equivalent width,
in that objects with large EW generally present low $L_{\rm [CII]}$/SFR ratios.
The results of Harikane et al. (2020) span the interval 
6$\lesssim$ log($L_{\rm [CII]}$/SFR~[$L_{\odot} M_{\odot}^{-1} yr$]) $\lesssim$ 7.5.
If we take at face value our estimates
for $L_{\rm [CII]}$ and SFR,  we obtain for this ratio the extreme value of log($L_{\rm [CII]}$/SFR)=5.26 $L_{\odot} M_{\odot}^{-1} yr$ which,
with plausible values for the Lyman-$\alpha$ EW $>120 \pm 8$ \AA (V19), confirms the observed anticorrelation between
[CII] deficit and Ly$\alpha$ EW (see also Harikane et al. 2018; Schaerer et al. 2020).\\
More recently, within the ALPINE-ALMA survey, 
Schaerer et al. (2020) presented an analysis of the $L_{\rm [CII]}$-SFR relation in a sample of $118$ star-forming galaxies at $z\sim~5$.
Their analysis indicates a substantially good agreement between the properties of their sample and the $L_{\rm [CII]}$-SFR relation in galaxies
from other compilations both at comparable or higher $z$ and in the local Universe.
Their study of an homogeneous sample spanning a cosmic interval of $\sim~13$ Gyr 
suggests no evolution, which points towards a universal $L_{\rm [CII]}$-SFR relation in normal star forming galaxies (with $SFR\ge 1 M_{\odot}/yr$),
with a behaviour similar to that shown by the collection of data in Fig.~\ref{fig_cii}.\\
In general, the  $L_{\rm [CII]}$-SFR relation at high redshift is poorly known in the faintest $L_{\rm [CII]}$ regime.  
In this context, the current study is of particular interest, as our investigation of the properties of
[CII] emission allows us to access a practically unexplored region of the $L_{\rm [CII]}$ - SFR plot. 
The sources that show the largest offset from the local relation are the two faint lensed galaxies
of Knudsen et al. (2016, yellow diamonds) and the lowest SFR-upper limit from Laporte et al. (2019), 
although in a recent reanalysis of ALMA data, Carniani et al. (2020) report a tentative [CII] detection for some of these galaxies,
possibly alleviating the discrepancy with the local $L_{\rm [CII]}$ - SFR relation. 

Of these star-forming galaxies at $z>6$, one galaxy (A383-5.1) presents SFR and $L_{\rm [CII]}$ values marginally compatible with the local relation
(grey area in Fig.~\ref{fig_cii}), whereas the lower point is for MS0451-H, a system with an extremely large
magnification  ($\mu=100$), which lies $\sim~1$~dex lower than the local relation. 
Taking the SFR measure of D1 at face value, our detection reveals an even larger offset with respect to the local relation.
However, a caveat is in order regarding large uncertainties on the measured SFR as derived from the SED fitting.
At this redshift, key SFR indicators like the rest-frame optical Balmer lines (e.g., H$\alpha$) will be observable with the NIRSPEC instrument
onboard the James Webb Space Telescope.

\subsubsection{Theoretical interpretation of the weak [CII] emissivity of high-z galaxies}
Previous theorethical studies have outlined that the
[CII] emission originates from the cold (with temperatures of a few 100 K) neutral medium and from photo-dissociation regions 
(PDR, Vallini et al. 2013). 
This seems to suggest that  its presence closely traces star formation
sites, resulting in a linear relation, as found by De Looze et
al. (2014) and Herrera-Camus et al (2015). While at low-redshift and
close to solar metallicity such a relation is well established, as
shown by several observations (DeLooze et al. 2014, Herrera-Camus et al. 2018) and also
numerical simulations (see, e.g. Lupi and Bovino 2020), significant
deviations can arise in different ISM conditions, like at lower
metallicity or in presence of a strong ionisations field, that are more typically
found in the high-redshift Universe.  To address the impact of such
conditions, several studies have analysed the [CII] emission from
typical high-redshift galaxies, by post-processing hydrodynamic
zoom-in cosmological simulations with \textsc{cloudy} (Ferland et al. 2017; see, e.g. Olsen et
al. 2017; Pallottini et al. 2017, 2019; Katz et al. 2019), or via 
ad-hoc methods, as in Arata et al. (2020), or also via on-the-fly
non-equilibrium chemistry (Lupi et al. 2020).  The main conclusion in
all these studies is that a [CII] deficit exists at high-redshift, 
most likely due to the starbursting nature of these galaxies
rather than their metallicity, since most of these systems are close to solar (see, e.g. Vallini et al. 2015, Lupi and Bovino 2020).
Other studies have evidenced a weak dependence of [CII] on metallicity. 
Harikane et al. (2020) showed that the $L_{[CII]}/SFR$ ratio does 
not show a strong dependence on metallicity, which to first approximation was interpreted as the result of the 
proportionality between C abundance and metallicity $Z$ and an inverse proportionality between
PDR column density and $Z$ in a dust-dominated shielding regime (Kaufman et al. 2006).
In this framework, if PDR give a large conribution to the [CII] emissivity, 
the [CII] luminosity is not expected to strongly depend on Z (see also Pallottini et al. 2019; Ferrara et al. 2019). 

An additional effect upon which a 
consensus is still missing is the cosmic microwave background (CMB), hotter at $z\sim 6$, 
that alters the excitation level population of the C+ atom and
provides a background against which the signal is observed. While some
studies concluded that such an effect can be important at 
high-redshift (Da Cunha et al. 2013, Vallini et al. 2015),
suppressing the emission from the intermediate- or low-density gas (i.e. with particle density values $n<10\rm\, cm^{-3}$), recent
numerical simulations have shown that, because of the typically higher
densities found in the $z\sim 6$ ISM, the actual effect of the CMB is 
mild (Lagache et al. 2018; Kohandel et al. 2019; Arata et al. 2020; Lupi et al. 2020). 

Taken at face value, our results further confirm the extremely weak emissivity of high-redshift systems, even though 
at present, theoretical attempts to model [CII] emission in objects with stellar masses as low as D1, or with a suitable resolution
to identify individual star clusters,  are still missing. \\ 
In the case of D1, the first, most intuitive reason for its low emissivity is the likely low metallicity of our system.
However, a low [CII] emissivity can also be imputable to other reasons, including low gas density (Vallini et al. 2015) 
and a strong radiation field, in which production of $C^{++}$ would be enhanced (e. g., Knudsen et al. 2016; Lagache et al. 2018; Harikane et al. 2020). 
A further explanation is the saturation of the emitted [CII] flux in a region of high SFR density
($\Sigma_{\rm SFR}=25 ~M_{\odot} kpc^{-2} yr^{-1}$) such as D1 (Ferrara et al. 2019). 

As for the metal content of D1 in the form of dust grains, more constrains can be achieved from the analysis of
its non-detection
in the continuum.

\subsubsection{Constraints on the dust content of D1 from the continuum flux}

In principle, even if a single measurement of the continuum emission at 
$\sim 1$ mm wavelengths (observer frame) does not allow us to constrain
the IR luminosity of a given system,  
under certain assumptions as discussed later, it can allow one to estimate the dust mass (Carilli \& Walter 2013).  

The possibility of constraining the dust content of a system like D1
is worthwhile, for at least two reasons. 
First, since dust grains
are important sinks of heavy elements, a derivation of the dust mass
allows us to derive constraints on the metallicity of our system.
Second, the origin of dust in low-metallicity, low-mass galaxies
is poorly known in general, and documented by a very limited amount of studies even in the local Universe 
(Fisher et al. 2014; Hunt et al. 2014, Lianou et al. 2019). \\
Several previous ALMA programs aimed at detecting the dust emission of various systems at high redshift by means
of observations at $\sim1$mm rest-frame (e. g., Venemans et al. 2018).
However, it was not uncommon that the dust  continuum
was not detected, especially at $z\sim 6$ (e. g. Aravena et al. 2016). 
When this occurs, the measured continuum sensitivity can usually be used to achieve bounds on the dust mass.\\
Also in the case of D1, no detection was found from the analysis of the continuum map, but 
a $3\sigma$ upper limit on its dust mass was derived from the sensitivity as described below.

The FIR continuum of star-forming systems is generally dominated by dust emission and its spectrum 
is assumed to follow a modified blackbody (e. g., Priddey \& McMahon 2001, Bianchi 2013, Gilli et al. 2014). 
Assuming single temperature dust, in an optically thin regime 
 an upper limit on the dust mass can be computed as:
\begin{equation}
M_{\rm  dust} \le  \frac{D^2_L~S_{\nu}}{(1+z)~k_{\rm  \nu}(\beta)B_{\rm  \nu}(T)}~ M_{\odot}/beam, 
\end{equation}
where 
$k_{\rm  \nu}(\beta)=0.77 (\nu/352 GHz)^\beta$ cm$^2$~g$^{-1}$
is the dust mass opacity coefficient (Dunne et al. 2000) and 
$B_{\rm  \nu}(T)$ is the Planck function.
In our case $S_{\nu}=3~\sigma_{\rm  cont}$ and we assume that the system is point-like, i.e. $N_{\rm beam}=1$. 

Two further assumptions need to be made on the spectral 
index $\beta$ and on the dust temperature $T$. 
These quantities are poorly known in systems like D1, 
hence here 
we test two different possibilities.
In one case, we regard the well-known Himiko system (Ouchi et al. 2013)
as the prototype of a star-forming system at $z>6$, and assume the same
parameters adopted for this system to derive an upper limit on its dust mass, i. e. 
$\beta=1.5$ and $T=40$ K.
This choice is motivated by the fact that, despite Himiko appears as a vigorous star-forming system,
it is also dust-poor and, for this reason, it can be regarded as an 
analogues of a metal-poor system, comparable to a few local blue compact dwarf galaxies (Fisher et al. 2014) 
and possibly like the subject of our study. \\
In the second case, we consider the recent study of Inoue et al. (2020),
who developed an 
algorithm to solve the radiative-transfer equation and 
derived the temperature and mass of dust in high-$z$ galaxies,
assuming dust to be in radiative equilibrium and considering a set of different geometries.
Assuming $\beta=2$, Inoue et al. (2020) estimated dust temperatures between 38 K and 70 K.
A similar result has been also found by Sommovigo et  al. (2020) who developed a model for dust temperature within giant molecular clouds (GMC). 
These authors found that, due to the smaller size and higher turbulent nature of GMC at high-z, 
the dust is warmer that in local ones and can reach temperature values as high as T=60 K.

\begin{figure*}
\includegraphics[height=11cm, width=11cm]{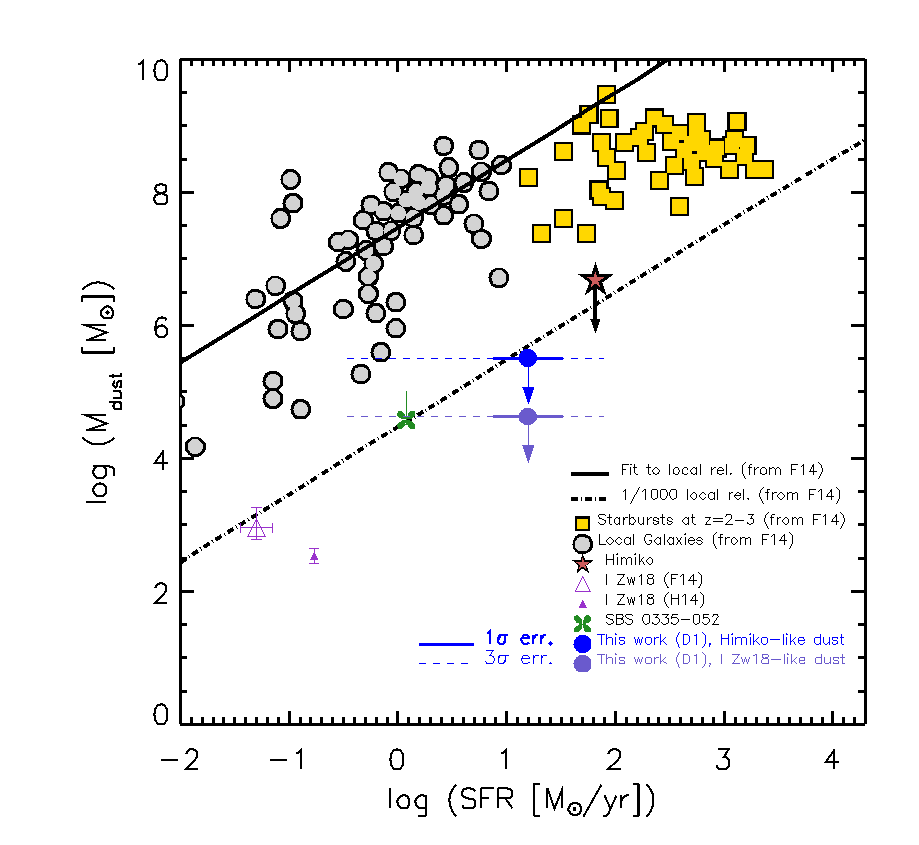}
\caption{Observed relation between dust mass and star formation rate as obtained in the present work from our non-detection in the
continuum, compared to the values derived in local and distant star-forming galaxies.
Our estimates are the blue solid and violet circles with horizontal error bars and downpointing arrows, obtained if Himiko-like
dust grains (charatectised by spectral index $\beta=1.5$ and dust temperature $T=40$ K as in Ouchi et al. 2013)
and I Zw18-like (with spectral index $\beta=2$ and dust temperature 70 K as in Hunt et al. 2014) dust are assumed,
respectively.
Both our upper limits are reported  with their 
  $1\sigma$ (blue and violet thick solid lines) and $3\sigma$ (blue and violet thin dashed lines) error bars. 
The gray solid circles and yellow squares are compilation of
local star forming galaxies and starbursts at $z=2-3$, respectively, from Fisher et al. (2014).
The light red star is the upper limit on the dust mass of Himiko as derived in Fisher et al. (2014). 
The solid and dot-dashed black lines are a linear fit to the local relation from Fisher et al. (2014) and 
the same fit rescaled down by a factor 1000, respectively. 
The open and solid purple triangles with error bars are  dust mass estimates for I Zw18 presented by
Fisher et al. (2014) and Hunt et al. (2014), respectively. The green clover is the value estimated by
Hunt et al. (2014) for the local low-metallicity dwarf galaxy SBS 0335-052, where the error bar for the dust mass considers 
the upper limit of $10^5~M_{\rm  \odot}$ from Cormier et al. (2017).} 
\label{fig_dust}
\end{figure*}

It is worth noting that a type of dust with $\beta=2$ and $T=70$ K is similar to the result of Hunt et al. (2014)
obtained in their single-temperature blackbody fits to the spectrum of I Zw18 as observed with ALMA (for this reason
the values obtained in this case are labelled 'I Zw18-like dust' in Fig.~\ref{fig_dust}) .\\

The upper limits on the dust mass of D1 obtained as described in this section are reported
in the $M_{\rm  dust}$-SFR relation shown in Fig.~\ref{fig_dust}, compared with other estimates
from the literature in local and distant galaxies.
The compilation of data in local and distant (i.e. at $z\sim2-3$) galaxies shown in Fig.~\ref{fig_dust} is 
taken from Fisher et al. (2014). Also an empirical relation derived in local galaxies (see also Da Cunha et al. 2010)
is included, together with the upper limit obtained by Ouchi et al. (2013) in Himiko. \\
Other estimates of the dust mass in local low-metallicity dwarf galaxies (I Zw and SBS 0335-052)
are reported (for further details see caption of Fig.~\ref{fig_dust}).
The $M_{\rm  dust}$-SFR relation observed in local metal-poor galaxies deviates by more than 2 orders
of magnitude from the one observed in normal star-forming galaxies, which is a tight,
quasi-linear relation, explained by multiple physical processes which influence the star formation history of
galaxies such as starbursts, quenching and merging episodes (Hjorth et al. 2014; Calura et al. 2017).
The upper limit as derived by Ouchi et al. (2013) in Himiko at $z\sim 6.6$ indicates a
$M_{\rm  dust}$/SFR ratio similar to local metal-poor dwarfs.\\ 
The limits estimated for D1 in all the cases shown in Fig. ~\ref{fig_dust} indicate values by at least one order 
of magnitude lower than local normal galaxies and starbursts.
In particular, in the likely case of hot dust (T=70 K), our limit is consistent with the value
measured in one local metal-poor starburst galaxy, i. e. SBS 0335-052 (Hunt et al. 2014; Cormier et al. 2017).

This suggests that D1 might be characterised by an extremely low metal and dust content. 
This is compatible also with other facts, i. e. the low [CII] emissivity discussed above, as well
as the very blue slope of its ultraviolet spectrum. 

The production of dust in galaxies at epochs near reionization is a poorly known matter.
The appearence of dust is probably a rapid process, as galaxies at larger redshifts than Himiko and with comparable SFR values
show much larger dust masses (Mattsson et al. 2015; Hashimoto et al. 2018). 
More generally, very little is known on dust production at low metallicity and in galaxies with low SFR at both low and high redshift,
even in SFR regimes where tight, robustly constrained scaling relations followed by luminous galaxies are generally well established
(e. g., Da Cunha et al. 2010; Hjorth et al. 2014). 
This outlines the need of sensitive instruments which can allow us to trace the dust content either in local low metallicity
systems and in high redshift galaxies.
Of these instruments, the best candidate in the future is the infrared telescope SPICA,
which will reach unprecedented sensitivity at mid-IR and and  FIR wavelengths (e. g., Fern{\'a}ndez-Ontiveros, et al. 2017). \\

\subsubsection{On the presence of cold gas in proto-globular clusters}
D1 contains at least one very compact young stellar cluster,
as revealed by the strongly concentrated emission detected in its centre, 
with other, similar systems in its proximity. 
A fundamental question regarding the origin of high-$z$ star clusters and their link with
GCs concern their capability to retain their gas during and immediately after an episode of star formation 
and how much time they spend embedded in their natal gas cloud. 

In some cases, high-spatial resolution ALMA observations of 
young massive clusters (YMC) in nearby star-forming galaxies show that at ages of $\sim$ a few Myr, they are still embedded
in their parent molecular cloud, and that the cocoons of cold gas enshrouding the cluster are essentially
undisturbed (Turner et al. 2017; Silich et al. 2020). 

Other observations of YMC with stellar masses of $10^5 - 10^6 M_{\rm  \odot}$ in nearby dwarf, spiral 
and interacting galaxies indicate a small content or absence of cold gas in or nearby the clusters 
already a few Myr after their formation (Bastian et al. 2014; Cabrera-Ziri et al. 2015).
These evidences stress that the capability of young clusters to retain cold gas may depend on several factors,
including mass, size, age of the cluster and star formation efficiency (Silich \& Tenorio-Tagle 2017). 

Simulations of isolated compact stellar systems with sub-pc resolution, currently out of reach in a cosmological context, 
are suitable to resolve the internal
structure of bound clusters and the role of crucial phenomena, such as stellar feedback and 
gas accretion. 
These simulations have shown that the energetic feedback 
of massive stars, through both stellar winds and core-collapse supernovae,  
can blow out the dense gas in which 
they are embedded in a short time, of the order of $\sim 10$ Myr
(e. g., Calura et al. 2015; see also Silich \& Tenorio-Tagle 2017; Yadav et al. 2017),  
comparable to the estimated age of the stellar populations of D1 (V19).

Current scenarios for the formation of multiple stellar populations in GCs envisage
the presence of reservoirs of cold gas available nearby young clusters, ready to be accreted
after massive stars have stopped restoring energy into their surroundings, i. e. on timescales of 
the order of $\sim~30$ Myr (e. g., Naiman et al. 2011; D'Ercole et al. 2016; Calura et al. 2019).
These reservoirs of cold gas can be present if star clusters forms within the disc of a dwarf galaxy 
(e. g. Elmegreen et al. 2012; Kravtsov \& Gnedin 2005; Kruijssen 2015) or in mergers.  

In cosmological hydrodynamical simulations performed at the highest resolution,  
bound and dense clumps in gas-rich high-redshift galaxies have been identified with sizes and masses
marginally consistent with the typical values of star clusters (Ma et al. 2020; Phipps et al. 2020).
The birth of these clumps occurs in high-pressure regions with typical surface density values $\Sigma_{\rm g}>10^2~ M_{\odot}~pc^{-2}$,
namely in regions which are rich in molecular gas and where the formation of bound clusters is particularly
efficient (e. g., Li et at al. 2019), in conditions which may be particularly common in high-redshift galaxies,
promoted by the highest frequency of starbursts and interactions (Elmegreen \& Efremov 1997).  

Finally, we use the value obtained here for the [CII] luminosity of D1 to derive constraints on its molecular gas content. 
Recently, Zanella et al. (2018) 
presented a tight correlation between the [CII] luminosity and the molecular gas mass in main sequence and starburst galaxies.
This relation seems to hold across a wide range of SFR and metallicity, in particular down to
values 12 + log O/H$\sim 8$ or less, which may be typical for dwarf galaxies or metal-poor systems at hight redshift. 
Assuming a conversion factor between $L_{\rm [CII]}$ and $M_{\rm mol}$ of $\alpha_{\rm [CII]} = 31~M_{\odot}/L_{\odot} $, 
with our $L_{\rm [CII], D1}$ value from our tentative detection, we obtain a molecular gas mass $M_{\rm mol}=9 \times 10^7~M_{\odot}$.
This implies an upper limit on the dust-to-gas ratio of 0.0036 (computed for a dust mass $M_d<3.2 \times 10^5~M_{\odot}$) and might be indicative of the presence of 
a large reservoir of molecular gas in proximity of D1. 
This particular aspect needs to be investigated further in the next future, when deeper observations will be carried on\footnote{By means of the online ALMA
sensitivity calculator, we estimated that $\sim 10$ h integration will be sufficient to 
confirm our detection at 8 $\sigma$. }
and more contraints on the gas content of proto-GCs will be available.

\section{Conclusions} \label{sec4}

The properties of [CII] emitting galaxies are generally studied through the $L_{\rm [CII]}$-SFR relation,
which at high redshift was studied so far mostly in normal galaxies, i. e. generally with $L_{\rm [CII]}>10^7~L_{\odot}$. 
At present, very little about the behaviour of such relation is known in systems with lower [CII] luminosity and
very low $(\sim 10^7 M_{\odot})$ stellar mass.  
In this work we reported on recent ALMA observations of D1, a strongly magnified $M_{\rm  *} \sim 10^7 M_{\odot}$ system at $z\sim 6.145$
which contains globular cluster precursors, aimed at the detection of the [CII] emission line.
Since the discovery of GC progenitors at high redshift, 
ours is the first attempt to probe directly the physical properties of their cold gas through IR observations.
Our observations allowed us to constrain the [CII] emissivity of D1, with important consequences
for studies of extremely faint systems. 
After a careful analysis of the ALMA cube performed with an automatic procedure, 
testing various possible values for the linewidth value and searching for
features present in different regions of the ALMA FoV, 
we were able to find a $4\sigma$ tentative detection of [CII] emission, 
with a spatial offset from the centre of the
UV detection of $\sim$ 0.15\arcsec-0.2\arcsec,  
velocity offset with respect to the Ly$\alpha$ emission of +21 $\pm$ 128 km/s, linewidth of 110 $\pm$ 50 km/s 
and a flux density of 0.46 $\pm$ 0.22 mJy. This translates into an intrinsic (i. e. corrected for lensing)  
[CII] luminosity of $L_{\rm [CII]}=(2.9 \pm 1.4)~10^6 L_{\odot}$. 

Despite a large uncertainty on the SFR value, 
our estimate occupies a poorly explored region 
of the $L_{\rm [CII]}-SFR$ plot.
Taking the measured SFR at face value, our estimates
lie more than $\sim 1$ dex below the $L_{\rm [CII]}-SFR$ relation observed in local and high redshift systems. 

Our tentative detection confirms a deficiency of [CII] emission in high-redshift systems. 
To explain this finding we considered several possibilities, including a low-density gas 
driven by intense stellar feedback occurring in compact stellar clusters, a strong radiation field 
or a low metal content.
The later possibility appears confirmed by the upper limit on the dust mass as derived 
from the continuum flux, which for a reasonable range of the main parameters determining the dust mass (the FIR
spectral index $\beta$ and the dust temperature) indicate lensing-corrected dust masses $\le 3 \times 10^5~M_{\odot}$
and, in the likely case of hot dust (T=70 K), we find an upper limit of
$4 \times 10^4 M_{\odot}$, 
consistent with the values measured
in the local low-metallicity starburst SBS 0335-052. 
A low dust content is confirmed also by the extremely blue slope of its UV continuum (V19). \\
New GC precursors are continuosly discovered in lensed fields (Vanzella et al. 2020a; Kikuchihara et al. 2020), 
In the future, further deep ALMA pointings on some of these systems are needed 
to make progress in this relatively young field  
and improve our current knowledge of their physical properties.
As for D1, in order to better constrain both its [CII] emissivity and IR continuum and confirm our detection, deeper observations will be required, 
which will be proposed in next ALMA cycles.

\section*{Data Availability}
The data underlying this article
will become public on 2020-09-04, and will be available for 
download at the ALMA Science Archive at https://almascience.eso.org/asax/.

\section*{Acknowledgements}
This paper makes use of the following ALMA data: ADS/JAO.ALMA\#2018.1.00781.S. ALMA is a partnership of ESO (representing its member states), NSF (USA) and NINS (Japan),
together with NRC (Canada), MOST and ASIAA (Taiwan), and KASI (Republic of Korea),
in cooperation with the Republic of Chile. The Joint ALMA Observatory is operated by ESO, AUI/NRAO and NAOJ. 
An anonymous referee is acknowledged for several useful suggestions. 
We wish to  thanks L. Vallini and E. Vesperini for several interesting discussions. 
FC acknowledges support from grant PRIN MIUR 20173ML3WW$\_$001. 
EV, MM, PR and PB acknowledge support from PRIN MIUR 2017WSCC32. 
SC and AL acknowledge support from the ERC Advanced Grant INTERSTELLAR H2020/740120. 
We acknowledge funding from the INAF main-stream (1.05.01.86.31)








\appendix

\section{[CII] emissivity in Mock ALMA observations of D1}
\label{sec_mock}
In this appendix we discuss further the robustness of our estimate on the
[CII] luminosity of D1, considering the effects of
various parameters on the estimated SNR. 
First, we investigate 
the possibility that a faint [CII] emission might be result of flux
distributed over a large area, thus lowering the surface brightness of the source and leading to an underestimation of
its total luminosity (see Carniani et al. 2020).

By using mock ALMA observations, here we 
analyze how D1 would appear considering various assumptions on its 
linewidth, geometry of the source and [CII] luminosity. 

We use the CASA {\it simobserve} task to generate  mock observations of D1 by
setting the same antenna array configurations, weather condition, and
on-source exposure times of the real observations (see Sec. \ref{sec_obs}).
As input model we adopt the rest-frame UV
HST images of D1 and we generate synthetic observations for different values of [CII]
luminosity  ($6 \le Log(L_{\rm [CII]}/L_\odot\le 6.8$) and for three different line width values
of the [CII] profile, namely 50, 100, and 350 km/s. For each set of
parameters, we compute 10 ALMA simulations of D1. In each mock image we
estimate the SNR of the emission at the position of the artificial source, placed at the centre of the
field of view. 

\begin{figure*}
\includegraphics[height=7cm, width=15cm]{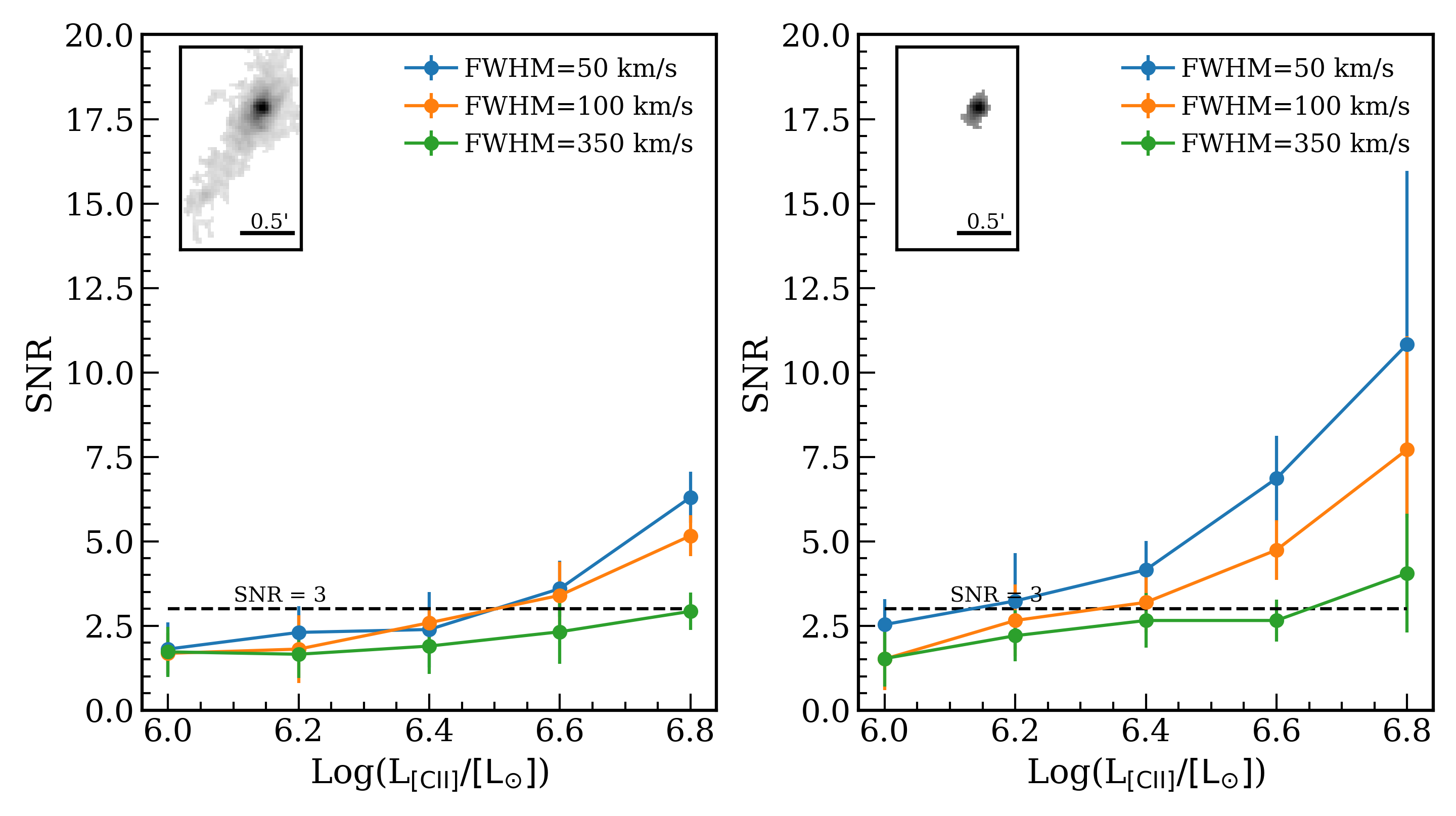}
\caption{Mock ALMA observations of D1 for various assumptions on the line width, on the luminosity and spatial distribution of the [CII] emission. 
In the left and right panels D1 is modelled as an extended and point-like source, respectively,
with the adopted light distribution reported in the top-left inset of each figure.
The plots show the SNR as a function of [CII] luminosity, computed assuming a
linewidth of 50 km/s (blue solid lines and solid circles), 100 km/s (orange lines)  and 350 km/s (green lines). 
For further details, see Appendix \ref{sec_mock}. 
}
\label{mock_obs}
\end{figure*}

Figure \ref{mock_obs} shows the SNR of the mock D1 images as a function of [CII]
luminosity and for different values of line width, expressed as full-width half maximum (FWHM).
The Figure shows two different assumptions on the distribution of the emission, with an
extended emission in the left panel and a point-like one on the right (the assumptions
on the geometry of the source are shown in the top-left inset of both panels). 

In the case of the extended source we considered all the emission down to flux values  
corresponding to 0.1 of the peak.  
Instead, in the case of the compact source, we performed a cut at flux values of 0.4 of the emission peak. 

In the case of extended (point-like) emission, we measure 
a SNR$>3$ for $L_{\rm [CII]} > 3.16\times10^6~L_{\odot}$ ($L_{\rm [CII]} > 1.8\times10^6~L_{\odot}$)
and FWHM=50-100 km/s. On the other
hand, if the line has a FWHM=350 km/s, it will be detected only when
$L_{\rm [CII]} >  4.0\times10^6~L_{\odot}$ in the case of point-like emission. 

A few more considerations are in order regarding the values obtained as a function of the assumed distribution of the emission, shown  
in the left panel of Fig. \ref{mock_obs}, which can be higher than the value estimated in Sec.~\ref{sec_lcii}. 
This discrepancy is 
associated to the assumption on the extension of the [CII] 
emission. While in Sec.~\ref{sec_lcii} we assumed that the [CII] emission
arises from the luminous knot of D1 that is spatially unresolved in 
current ALMA observations, in the simulations we consider an extended UV emission elongated towards south-west. In the latter case, 
the [CII] emission is spatially resolved out and part of the flux will be lost. 
From the comparison of the output luminosities estimated from the simulations with the input luminosities, 
using a circular aperture of radius 1\arcsec, we estimate that 
in spatially resolved observations the [CII] luminosity can be 
underestimated by a factor 1.3 (see also Carniani et al. 2020). 



\end{document}